\documentclass[dvips]{article}
\usepackage{icrctc07,subfigure}

\title{Energy cross-calibration from the first CREAM flight: transition radiation detector versus calorimeter}
\shorttitle{CREAM energy cross-calibration}
\authors{
P. Maestro$^7$, 
H.S. Ahn$^2$, 
P.S. Allison$^5$, 
M.G. Bagliesi$^7$, 
J.J. Beatty $^5$, 
G. Bigongiari$^7$,
P.J. Boyle$^3$, 
T.J. Brandt $^5$, 
J.T. Childers$^6$, 
N.B. Conklin$^4$, 
S. Coutu$^4$, 
M.A. DuVernois$^6$,
O. Ganel$^2$, 
J.H. Han$^8$, 
H.J. Hyun$^8$, 
J.A. Jeon$^8$, 
K.C. Kim$^2$, 
J.K. Lee$^8$, 
M.H. Lee$^2$, 
L. Lutz$^2$, 
P.S.~Marrocchesi$^7$,
A. Malinine$^2$,
S. Minnick$^9$, 
S.I. Mognet$^4$, 
S. Nam$^8$, 
S. Nutter$^{10}$, 
H. Park$^{11}$, 
I.H. Park$^8$, 
N.H. Park$^8$, 
E.S. Seo$^{1,2}$, 
R. Sina$^2$, 
S. Swordy$^3$, 
S.P. Wakely$^3$,
J. Wu$^2$, 
J. Yang$^8$,
Y.S. Yoon$^{1,2}$, 
R. Zei$^7$, 
S.Y. Zinn$^2$
}
\shortauthors{P. Maestro and et al}
\afiliations{
$^1$ Dept. of Physics, University of Maryland, College Park, MD 20742 USA \\
$^2$ Inst. for Phys. Sci. and Tech., University of Maryland, College Park, MD 20742 USA \\
$^3$Enrico Fermi Institute and Dept. of Physics, University of Chicago, Chicago, IL 60637, USA \\
$^4$Dept. of Physics, Penn State University, University Park, PA 16802, USA \\
$^5$Dept. of Physics, Ohio State University, Columbus, Ohio 43210, USA \\
$^6$School of Physics and Astronomy, University of Minnesota, Minneapolis, MN 55455, USA \\
$^7$Dept. of Physics, University of Siena and INFN, Via Roma 56, 53100 Siena, Italy \\
$^8$Dept. of Physics, Ewha Womans University, Seoul, 120-750, Republic of Korea \\
$^9$Dept. of Physics, Kent State University Tuscarawas, New Philadelphia, OH 44663, USA \\
$^{10}$Dept. of Physics and Geology, Northern Kentucky University, Highland Heights, KY 41099, USA \\
$^{11}$Dept. of Physics, Kyungpook National University, Taegu, 702-701, Republic of Korea
}
\email{paolo.maestro@pi.infn.it}

\abstract{The Cosmic Ray Energetics And Mass (CREAM) balloon experiment had two successful
flights in 2004/05 and 2005/06. It was designed to perform energy measurements from
a few GeV up to 1000 TeV, taking advantage of different detection techniques. The
first instrument, CREAM-1, combined a transition radiation detector 
with a calorimeter to
provide independent energy measurements of cosmic-ray nuclei. Each detector was
calibrated with particle beams in a limited range of energies. In order to assess the
absolute energy scale of the instrument and to investigate the systematic
effects of each technique, a cross-calibration was performed by comparing the
two independent energy estimates on selected samples of oxygen and carbon nuclei.}

\begin{document}
\maketitle
\section{Introduction}
CREAM is a balloon-borne experiment designed to perform  
direct measurements of the energy spectra and  elemental composition of cosmic rays (CR)
up to the PeV scale. Two instruments, launched from McMurdo in 2004 
and 2005, flew over Antarctica for 42 and 28 days, respectively.
Both instruments achieved single-element discrimination 
by means of multiple measurements of the particle charge provided by 
a pixelated silicon charge detector (SCD), a segmented 
timing-based particle-charge detector (TCD) and a Cherenkov detector (CD). 
The particle energy was measured by a thin ionization calorimeter (CAL)
preceded by a graphite target. 
During the first flight, the payload was equipped 
with a Transition Radiation Detector (TRD), 
thus allowing redundant energy measurements.
A detailed description of the instrument can be found
elsewhere \cite{ref1}. 
In this paper, we present an analysis, based on the first flight data, 
that shows how it is possible to cross-calibrate the TRD and the calorimeter 
to assess the absolute scale of energy measurements in CREAM.
\section{Complementary techniques for particle energy measurement}
Direct measurements of charged CR are based on identification 
of the incoming particle and  measurement of its energy.
At present, the main active techniques for the determination of  CR energy at TeV scale 
are based on Ionization Calorimeters (IC) 
and TRDs. A combination of a IC and a TRD 
was implemented in the first CREAM payload.\\
The CREAM-1 TRD is made of 512 single-wire mylar thin-walled proportional tubes 
inserted in a polystyrene foam radiator structure and arranged in 8 layers,
 with alternating X/Y orientations. 
The 2 cm diameter tubes 
are filled with a mixture of 95\% xenon 5\% methane at 1 atm
which has a high efficiency for TR x-rays of a few tens of keV. 
The TRD can measure the energy of primary nuclei with Z$>$3 by multiple independent sampling of 
the energy deposit per unit pathlength (dE/dx) in the tubes. The ionization energy loss 
increases logarithmically  with the Lorentz factor $\gamma$ in the relativistic rise region which extends  from minimum ionization (MIP) 
to the Fermi plateau. In the case of Xe the ratio plateau/MIP is $\sim$ 1.5.   
At energies higher than a few hundred GeV/n, the ionization energy loss of a charged particle in Xe saturates. 
Nevertheless, the energy can be determined from the additional ionization produced in the tubes by the 
TR photons emitted as the particle crosses the foam radiator. 
A reliable estimate of the energy deposit 
requires a precise measurement of the  pathlength of the primary particle traversing the TRD.
For this purpose, the detector
has been designed to provide accurate particle tracking, with a resolution of 
the impinging point of the primary particle on the TCD to better than 2 mm. This allows to correct 
 the response of TCD and CD for 
spatial non uniformities; it is also essential to 
identify with a low probability of confusion, the TCD paddles and SCD pixel traversed by the 
primary particle and hence reconstruct its charge.
Although the main purpose of the CD 
is to provide, combined with the TCD signal, 
a trigger for relativistic high-Z nuclei, 
it can  also be used 
to measure the velocity of particles at low energies in the range 
from the Cherenkov threshold ($\gamma\sim1.35$) up to  saturation ($\gamma\sim10$).  
For a detailed description of the TRD and its performance 
during the flight 
see \cite{ref2}.\\
The CAL is a stack of 20 tungsten plates (50$\times$50  cm$^{2}$, 
each 1 X$_0$ thick) with interleaved active layers 
instrumented with 1 cm wide ribbons of 0.5 mm diameter scintillating fibers.
A 0.47 $\lambda_{int}$ thick carbon target preceding the calorimeter  induces a nuclear 
interaction of the primary particle which initiates a hadronic shower.    
The electromagnetic (e.m.) core of the shower is  imaged by the CAL which is  sufficiently thick   
to contain the shower maximum and finely grained to provide 
shower axis reconstruction. 
The resolution 
of the impact point 
on the SCD is about 1 cm.
The concept of 
IC is imposed by the requirement of 
weight reduction, making practically impossible to fly a conventional ``total containment'' hadronic calorimeters.
In a thin calorimeter, where only the e.m. core of the hadronic shower is sampled, the energy resolution 
is affected by the statistical fluctuations 
in the fraction of energy carried by $\pi^0$ secondaries
produced in the shower, whose decays generate the e.m. cascade. 
As a result, the energy resolution is poor by the standards of total containment hadron calorimetry
in experiments at accelerators. 
Nevertheless, it is sufficient to reconstruct the steep energy spectra of CR
nuclei with a nearly energy independent resolution.
\section{Calibrations with particle beams}
Both the TRD and the CAL were calibrated independently at CERN before the final integration 
in the payload. 
The CAL was tested  
with beams of protons, electrons and heavy ions.  
While protons and electrons were mainly used to equalize the single ribbons for non-uniformity in  
light output and gain differences among the photodetectors, 
a beam of  ion fragments 
was used to verify
the linear response of the CAL up to about 8.2 TeV and to measure 
a nearly flat resolution 
at energies above 1 TeV \cite{CALbeamtest}.
The TRD was tested  with protons, electrons and pions in a range of Lorentz factors
from $\sim$150 to 3$\times$10$^5$. This allowed to calibrate the instrument  
in two separate intervals  along the
  specific ionization curve: on the Fermi plateau and in the region of TR saturation. 
A MonteCarlo (MC) simulation of the apparatus based on GEANT4, including 
a modelization of the TR emission from the radiator, showed a remarkable agreement 
with the experimental data and was used to extend the calibration of 
the detector response
at lower $\gamma$ values than the ones available with the beam, i.e. 
to the relativistic rise region (10-500 GeV/n) \cite{Swordy}. However, an independent calibration based on flight data 
is preferable in order to validate the MC and to avoid systematic errors in the energy measurement of 
CR nuclei of a few hundred GeV/n. In fact, the TRD capability to provide a precise energy determination
in the relativistic rise region 
is essential for an accurate measurement
of the flux ratio of secondary to primary elements in CR, 
which is one of the main CREAM goals. 
\begin{figure}[h]
\begin{center}
\includegraphics[scale=0.425]{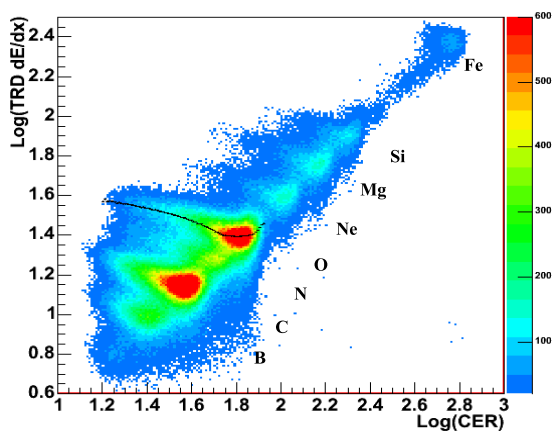}
\caption{Correlation of dE/dx measurements from the TRD and the CD signals, both expressed in arbitrary units,
for different nuclei populations from flight data. 
The black line is the average TRD response for O nuclei as a function of the CD signal.
}
\label{fig1}                            
\end{center}
\vspace{-0.4cm}
\end{figure}
\begin{figure*}
\begin{center}
\subfigure[]
{
\includegraphics[scale=0.35]{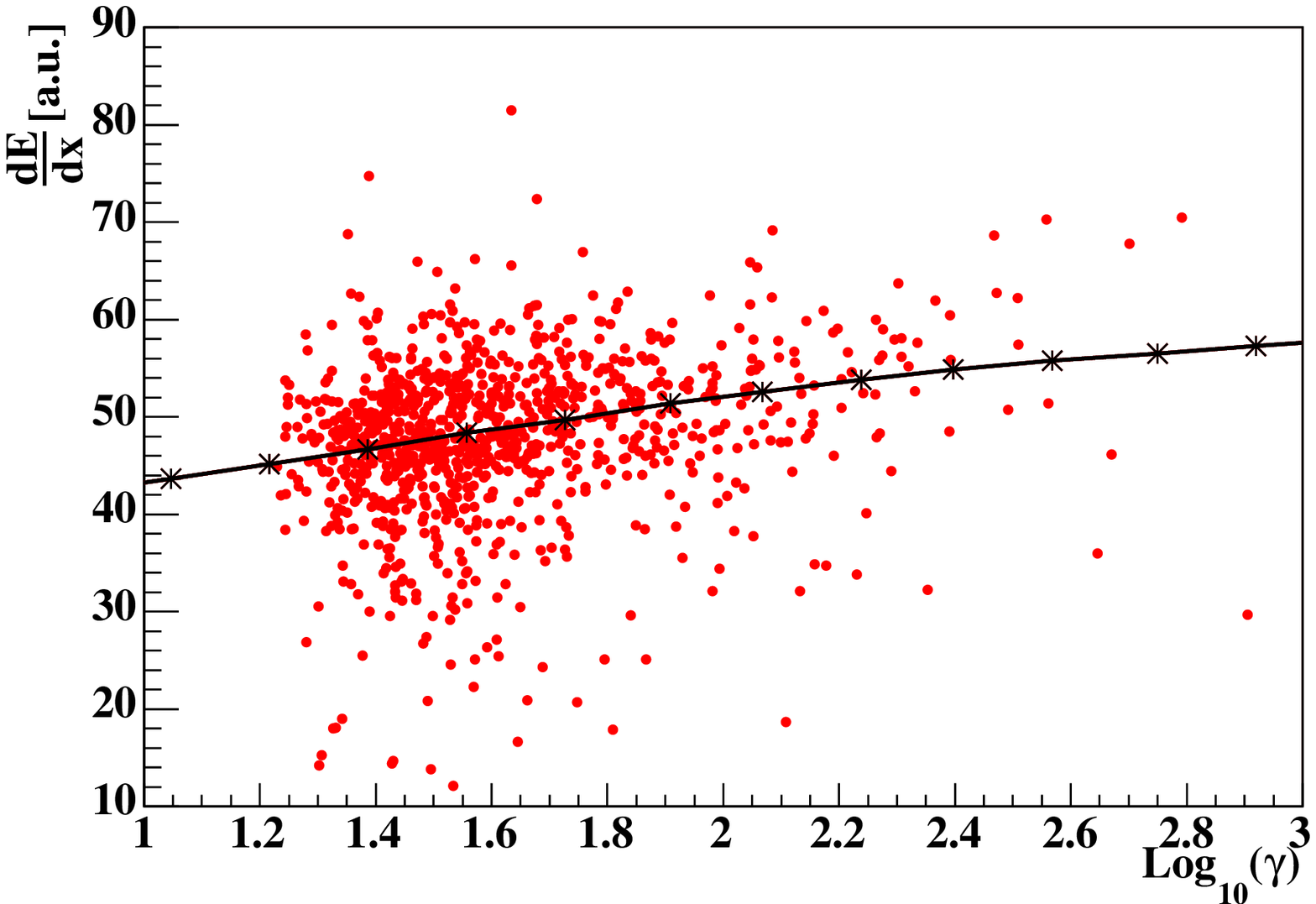}
\label{fig2a}                            
}
\subfigure[]
{
\includegraphics[scale=0.35]{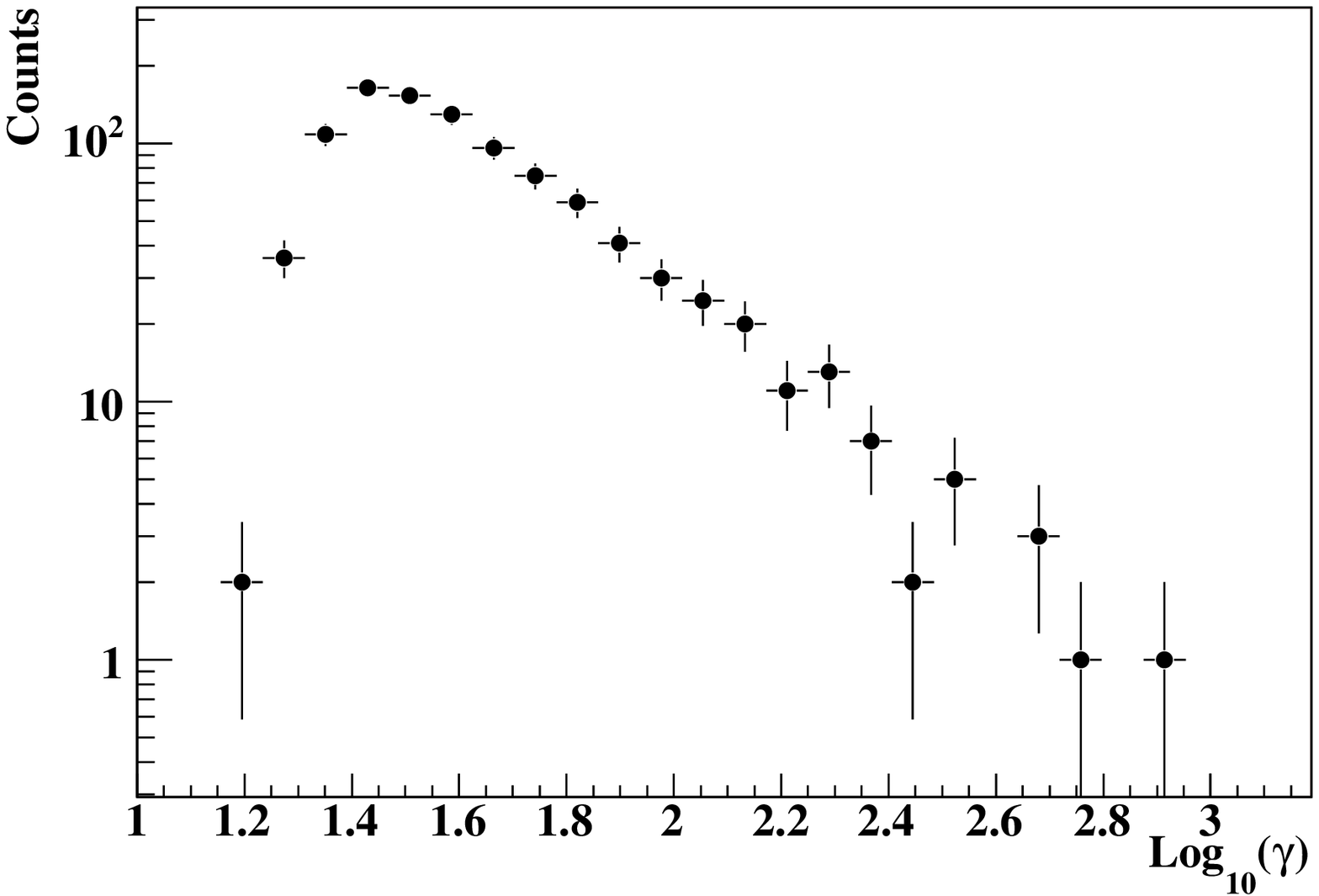}
\label{fig2b}                            
}
\vspace{-0.3cm}
\caption{(a) dE/dx measured with the TRD vs. the Lorentz factor $\gamma$ (in Log$_{10}$ scale), calculated from the energy deposit in the CAL, for the O sample from flight data. The superimposed curve is the GEANT4 prediction for the specific ionization in xenon.
(b) Distribution of the reconstructed $\gamma$ for the O  selection.}
\label{fig2}                            
\end{center}
\vspace{-0.5cm}
\end{figure*}
\section{Cross-calibration with flight data}
The TRD can be calibrated with flight data in energy intervals not covered at the beam test,
by correlating its response with the energy measurements provided by the CD and the CAL.
\begin{figure*}
\begin{center}
\hspace{-0.2cm}
\includegraphics[scale=0.5]{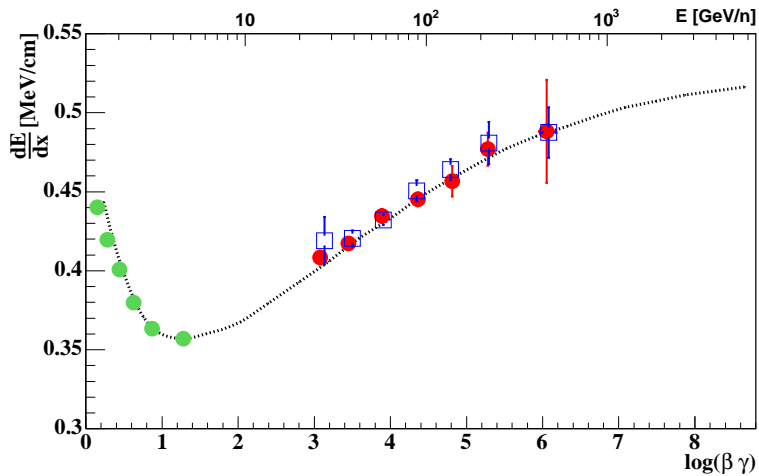}
\caption{TRD energy calibration with O (filled circles) and C (open squares) samples from flight data. 
The energy is measured with the CD below the minimum of ionization (green circles) and with the CAL 
in the relativistic rise region (red circles and blue squares). The dotted line 
represents the specific ionization curve in xenon predicted by GEANT4. 
}
\label{fig3}                            
\end{center}
\vspace{-0.5cm}
\end{figure*}
Events were selected by requiring that the primary particle track reconstructed by the TRD
was within the TCD acceptance and had at least four proportional tubes hit in each view.  
The pulse heights of the track-matched TCD paddles 
were combined with the CD signal to get a measurement of the particle charge.
An excellent separation of the charge peaks for elements from beryllium to silicon was obtained, 
with a charge resolution for carbon and oxygen better than 0.2 $e$ \cite{Coutu}.
The energy deposit per unit  pathlength (dE/dx) in the TRD was extracted with a likelihood fit, taking into account
the impact parameters of the primary particle track and the signal in each tube. 
Events were rejected if the two measurements of dE/dx, obtained by using independently the X and Y views of the TRD, 
disagreed by more than 20\%.
The correlation of the measured dE/dx and CD signal (Figure \ref{fig1})
allowed to calibrate the TRD response
in the region below the minimum of the specific ionization curve.
Six different intervals of $\gamma$ were selected with the CD,
and in each interval the average dE/dx was measured.
The scale factor to convert from arbitrary units (a.u.)
to MeV/cm was obtained 
by matching the minimum  ionization of O nuclei 
to the corresponding point of the MC simulated curve 
(Figure \ref{fig3}).
The Cherenkov emission yield saturates above $\gamma \sim$10, therefore the 
calibration  of the TRD in the relativistic rise region has to rely on the CAL energy measurement.
For this purpose, two samples of C and O nuclei were identified with the primary particle 
crossing the TRD and then generating a shower  in the CAL module. 
The dE/dx measurement was correlated with the particle 
energy measured  with the CAL. The scatter plot for the O sample is shown in Figure \ref{fig2} 
together with its projection on the horizontal axis, which represents 
the energy distribution reconstructed by the CAL. At values of Log$_{10} \gamma >$ 1.5, 
it exhibits the typical power-law behaviour expected from
the energy dependence of the differential cosmic-ray spectrum.
The range of measured $\gamma$ has been divided into 7 bins 
wherein the mean $\gamma$ and dE/dx values have been calculated. 
In this way the relativistic rise of the energy loss distribution was sampled as shown in Figure \ref{fig3}. 
The carbon points have been rescaled by taking into account the Z$^2$ dependence of dE/dx, 
in order to plot them on the same scale as the oxygen data. 
The TRD calibration based on the CAL energy measurement
shows  excellent agreement with the MC simulation. 
In this way, we proved that the GEANT4 prediction for the specific ionization in Xe can be
used as a reliable calibration to infer the primary particle energy from the dE/dx measured with the TRD, 
even at energies where the detector was not tested at accelerator beams. 
Moreover, the correct understanding of the absolute scale of the CAL energy measurement was confirmed.
\section{Conclusions}
A preliminary analysis of the data from the first flight of CREAM confirmed the possibility
to cross-calibrate the energy measurements of TRD and calorimeter.
\section{Acknowledgments}
This work is supported by NASA, NSF, INFN, PNRA, KICOS, MOST and CSBF. 

\begin{thebibliography}{99}
\bibitem{ref1} 
H.~S. Ahn et al. The Cosmic Ray Energetics and Mass (CREAM) Instrument. 
{\em In press in Nucl. Inst. Meth. A (2007)}
\bibitem{ref2} 
S.~P. Wakely et al.
{\em Adv. Sp. Res. (2007), doi:10.1016/j.asr.2007.03.080, in press}
\bibitem{CALbeamtest} 
H.~S. Ahn et al.
{\em Nucl. Phys. B (Proc. Suppl.) 150 (2006) 272-275}
\bibitem{Swordy} 
P.~J. Boyle, S.~P. Swordy and S.~P. Wakely.
{\em Proc. of 28$^{th}$ ICRC (2003) 2233}
\bibitem{Coutu} 
S. Coutu et al. 
{\em Nucl. Inst. Meth. A572 (2007) 485-487}
\end{thebibliography}

\end{document}